\documentclass{PoS}

\usepackage[utf8x]{inputenc}

\usepackage{amssymb}
\usepackage{amsmath}

\newcommand{\rt}{{\mathbf{r}}}
\newcommand{\xt}{{\mathbf{x}}}
\newcommand{\bt}{{\mathbf{b}}}

\newcommand{\yt}{{\mathbf{y}}}

\newcommand{\pt}{{\mathbf{p}}}
\newcommand{\qt}{{\mathbf{q}}}
\newcommand{\kt}{{\mathbf{k}}}

\newcommand{\ptt}{p_\perp} 
\newcommand{\ktt}{k_\perp} 
\newcommand{\qtt}{q_\perp} 
\newcommand{\ltt}{l_\perp} 
\newcommand{\lt}{\mathbf{l}}

\newcommand{\xif}{{\xi}_\text{f}}
\newcommand{\xf}{{x}_\text{f}}

\newcommand{\ud}{\mathrm{d}}

\newcommand{\tr}{\, \mathrm{Tr} \, }

\newcommand{\nc}{{N_\mathrm{c}}}

\newcommand{\cf}{C_\mathrm{F}}

\newcommand{\qs}{Q_\mathrm{s}}

\newcommand{\as}{\alpha_{\mathrm{s}}}

\newcommand{\ical}{\mathcal{I}}
\newcommand{\jcal}{\mathcal{J}}
\newcommand{\scal}{\mathcal{S}}

\def\figscale{1.35}

\title{Single inclusive hadron production in pA collisions at NLO}

\ShortTitle{Single inclusive hadron production in pA collisions at NLO}

\author{\speaker{B. Duclou\'e}\\
        Department of Physics, P.O. Box 35, 40014 University of Jyväskylä, Finland\\
        and \\
        Helsinki Institute of Physics, P.O. Box 64, 00014 University of Helsinki,
        Finland \\
        E-mail: \email{bertrand.b.ducloue@jyu.fi}}
    
\author{T. Lappi\\
        Department of Physics, P.O. Box 35, 40014 University of Jyväskylä, Finland\\
        and \\
        Helsinki Institute of Physics, P.O. Box 64, 00014 University of Helsinki,
        Finland \\
    	E-mail: \email{tuomas.v.v.lappi@jyu.fi}}
    
\author{Y. Zhu\\
        Department of Physics, P.O. Box 35, 40014 University of Jyväskylä, Finland\\
        and \\
        Helsinki Institute of Physics, P.O. Box 64, 00014 University of Helsinki,
        Finland \\
    	E-mail: \email{yan.zhu@jyu.fi}}

\abstract{We study single inclusive forward hadron production in high energy proton-nucleus collisions at next-to-leading order in the Color Glass Condensate framework. Recent studies have shown that the next-to-leading order corrections to this process are large and negative at large transverse momentum, leading to negative cross sections. We propose to overcome this difficulty by introducing an explicit rapidity factorization scale when subtracting the rapidity divergence into the evolution of the target.}

\FullConference{XXIV International Workshop on Deep-Inelastic Scattering and Related Subjects\\
		11-15 April, 2016\\
		DESY Hamburg, Germany}

\begin{document}

\section{Introduction}

High energy hadronic reactions, such as the ones performed at RHIC and the LHC, allow to access a region where gluon densities can be nonperturbatively large, even in the presence of a hard scale. This regime can be described by the Color Glass Condensate effective field theory, in which hadrons probed at small $x$ are described in terms of classical color fields. Of particular interest to study these dynamics are reactions where a dense hadron is probed by a simple dilute projectile, such as a proton at large $x$ which can be described in terms of well known collinear parton distributions functions.
One such process is the single inclusive forward hadron production in high energy proton-nucleus collisions, for which the expression for the cross section at leading order was derived in Ref.~\cite{Dumitru:2002qt}. Several works using this leading order formalism were able to provide a reasonable description of the trend of experimental data~\cite{Dumitru:2005gt,Albacete:2010bs,Tribedy:2011aa,Rezaeian:2012ye,Lappi:2013zma}. However, at this order the absolute normalization of the cross section is not well determined. Therefore it is important to extend this formalism to higher orders.
An important step in this direction was performed in Refs.~\cite{Chirilli:2011km,Chirilli:2012jd}, where the cross section for this process was computed at next-to-leading order. However the first numerical implementation of these expressions showed that at large transverse momenta the NLO corrections are negative and large enough to make the total cross section negative~\cite{Stasto:2013cha}. There have been several proposals to solve this problem, e.g. \cite{Kang:2014lha,Altinoluk:2014eka,Watanabe:2015tja}.
Here we propose an alternative way by introducing an explicit rapidity factorization scale when subtracting the rapidity divergence, similarly to what is done to absorb the collinear divergence in the DGLAP evolution of the parton distribution functions and fragmentations functions.

\section{Formalism}

Here we will only consider the $q \to q$ channel for simplicity. The expression for the NLO multiplicity in this channel can be obtained from Ref.~\cite{Chirilli:2012jd} after removing the integration over the impact parameter $\bt$:
\begin{align}\label{eq:nlosigma}
\frac{\ud N^{pA\to hX}}{\ud^2\pt \ud y_h}
=&
\int_\tau^1 \frac{\ud z}{z^2}D_{h/q}(z) x_p q(x_p) \frac{\scal^{(0)}(\ktt) }{(2\pi)^2}
\\ \nonumber
& +  \frac{\as}{2\pi^2} \int \frac{\ud z}{z^2}D_{h/q}(z) 
\int_{\tau/z}^1 \ud \xi \frac{1+\xi^2}{1-\xi}
\frac{x_p}{\xi} q\left(\frac{x_p}{\xi}\right) \left\{\cf \ical(\ktt,\xi) + \frac{\nc}{2}\jcal(\ktt,\xi) \right\}
\\ \nonumber
& - \frac{\as}{2\pi^2} \int \frac{\ud z}{z^2}D_{h/q}(z) 
\int_{0}^1 \ud \xi \frac{1+\xi^2}{1-\xi}
x_p q\left(x_p \right) \left\{\cf \ical_v(\ktt,\xi) + \frac{\nc}{2}\jcal_v(\ktt,\xi) \right\},
\end{align}
where 
\begin{align}\label{eq:J_Jv}
\ical(\ktt,\xi) \!\! =&
\int \frac{\ud^2 \qt}{(2\pi)^2} \scal(\qtt)
\left[\frac{\kt-\qt}{(\kt-\qt)^2} - \frac{\kt-\xi \qt}{(\kt-\xi \qt)^2} \right]^2
\\ \nonumber
\jcal(\ktt,\xi) \!\! =& \!\!
\int \frac{\ud^2 \qt}{(2\pi)^2} 
\frac{2(\kt-\xi\qt)\cdot(\kt-\qt)}{(\kt-\xi\qt)^2(\kt-\qt)^2}
\scal(\qtt)
\!-\!\!\!\int \frac{\ud^2 \qt}{(2\pi)^2} \frac{ \ud^2\lt}{(2\pi)^2}
\frac{2(\kt-\xi\qt)\cdot(\kt-\lt)}{(\kt-\xi\qt)^2(\kt-\lt)^2}
\scal(\qtt)\scal(\ltt)
\\ \nonumber
\ical_v(\ktt,\xi) \!\! =&
\scal(\ktt)\int \frac{ \ud^2 \qt }{(2\pi)^2}
\left[\frac{\kt-\qt}{(\kt-\qt)^2} - \frac{\xi\kt-\qt}{(\xi \kt-\qt)^2} \right]^2
\\ \nonumber
\jcal_v(\ktt,\xi) \!\! =&
\scal(\ktt)
\left[
\int \frac{\ud^2 \qt}{(2\pi)^2} 
\frac{2(\xi\kt-\qt)\cdot(\kt-\qt)}{(\xi\kt-\qt)^2(\kt-\qt)^2}
-\int \frac{\ud^2 \qt}{(2\pi)^2} \frac{  \ud^2\lt}{(2\pi)^2}
\frac{2(\xi\kt-\qt)\cdot(\lt-\qt)}{(\xi\kt-\qt)^2(\lt-\qt)^2}
\scal(\ltt)
\right].
\end{align}
The kinematical variables involved in these expressions are $\pt = z \kt$, $x_p = \ktt e^{y_h}/\sqrt{s}$, $\tau = z x_p$, $x_g = \ktt e^{-y_h}/\sqrt{s}$, $\ptt=|\pt|$, $\qtt=|\qt|$, $\ktt=|\kt|$, and $\ltt=|\lt|$. The additional variable appearing at next-to-leading order, $\xi$, is the longitudinal momentum fraction of the incoming quark taken by the fragmenting quark. The longitudinal momentum fraction carried by the radiated gluon is thus $1-\xi$, i.e.  $\xi \to 1$ corresponds to the limit of soft gluon emission.
Equations~(\ref{eq:nlosigma}) and~(\ref{eq:J_Jv}) are expressed as a function of $\scal$, which is the Fourier transform of the dipole operator in the fundamental representation: $\scal(\ktt)=\int \ud^2\rt e^{-i\kt\cdot\rt} S(\rt)$, with $S(\rt=\xt-\yt)=\left< \frac{1}{\nc}\tr U(\xt)U^\dag(\yt) \right>$.

Equations~(\ref{eq:nlosigma}) and~(\ref{eq:J_Jv}) are affected by two types of divergences, which have to be factorized in the evolution of nonperturbative quantities.
The first type of divergence is the collinear divergence. It affects only the NLO terms proportional to $\cf$. For these terms we use the same treatment as in Ref.~\cite{Chirilli:2012jd}: by using dimensional regularization, these divergences can be absorbed in the DGLAP evolution of the fragmentation functions $D_{h/q}(z)$ and quark PDFs $q(x)$.
The second type of divergence is the rapidity divergence, which affects NLO terms with a color factor $\nc/2$. One can see from Eq.~(\ref{eq:J_Jv}) that the transverse momentum integrals in $\jcal$ and $\jcal_v$ are finite but these terms do not vanish when $\xi \to 1$. Therefore they produce a divergence in this limit because of the factor $1/(1-\xi)$ appearing in Eq.~(\ref{eq:nlosigma}). This limit corresponds to soft gluon emission, thus it is natural to absorb this divergence in the evolution of the target. For this the renormalized $\scal$ was defined in Ref.~\cite{Chirilli:2012jd} as
\begin{equation}\label{eq:cxysub}
\scal(\ktt)  =  \scal^{(0)}(\ktt)
 + 2 \as \nc  \int_0^1 \frac{\ud \xi}{1-\xi} 
\left[\jcal(\ktt,1) - \jcal_v(\ktt,1)\right] ,
\end{equation}
which in coordinate space corresponds to an integral form of the Balitsky-Kovchegov equation~\cite{Balitsky:1995ub,Kovchegov:1999yj}.
This definition of the renormalized dipole cross section is however not unique. Indeed, one could subtract, instead of the integral over the whole $\xi$ interval, only contributions where $\xi$ is larger than a certain scale $\xif$. Thus we replace 0 by $\xif$ in the lower limit of the $\xi$ integral in Eq.~(\ref{eq:cxysub}), i.e. the original results of Ref.~\cite{Chirilli:2012jd} correspond to $\xif=0$.
By introducing $\xif$ we make the hard part explicitly dependent on this factorization scale. This dependence should cancel up to NLO accuracy with the corresponding dependence of the dipole cross section on the rapidity up to which it is evolved. This is similar to the way collinear divergences are absorbed in the DGLAP evolution of $D_{h/q}(z)$ and $q(x)$.

\section{Results}

In this section we demonstrate the importance of the choice of $\xif$. We will here consider the Golec-Biernat and Wüsthoff (GBW)~\cite{GolecBiernat:1998js} model for the dipole cross section. In this model both $S(\rt)$ and $\scal(\ktt)$ have simple gaussian expressions, enabling us to perform some of the integrals analytically:
\begin{equation}
S(\rt)=e^{-\rt^2 \qs^2/4} \, , \quad \scal(\ktt)=\frac{4\pi}{\qs^2}e^{-\ktt^2/\qs^2} , \quad \qs^2=c A^{1/3} Q_{s 0}^2 \left(\frac{x_0}{x}\right)^{\lambda},
\end{equation}
with $A$ being the atomic number of the target nucleus, $c=0.56$, $Q_{s 0}=1$ GeV, $x_0=3.04 \times 10^{-4}$ and $\lambda=0.288$~\cite{GolecBiernat:1998js}.
The expressions for the NLO cross section in this model were obtained in the large $\nc$ limit in Ref.~\cite{Chirilli:2012jd}. Here we use the corresponding expressions at finite $\nc$~\cite{Ducloue:2016shw} because we need to separate $\cf$-terms affected by the collinear divergence and $\nc$-terms affected by the rapidity divergence.
For the other parameters in our calculation we choose $\sqrt{s}=200$ GeV, $\alpha_s=0.2$, $\mu^2=10$ GeV$^2$ and $y_h=3.2$. We use the DSS~\cite{deFlorian:2007aj} and MSTW 2008~\cite{Martin:2009iq} NLO parametrizations for the fragmentation functions $D_{h/q}(z)$ and quark PDFs $q(x)$ respectively.

\begin{figure}
	\centering
	\includegraphics[scale=\figscale]{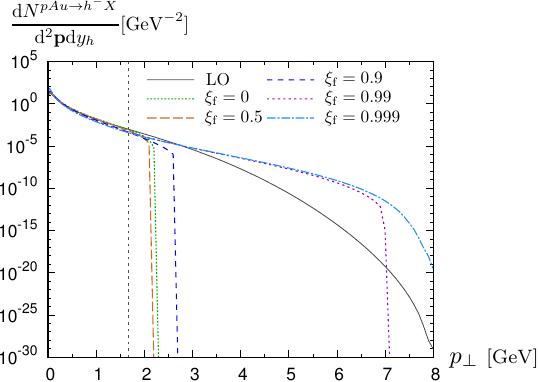}
	\hspace{0.3cm}
	\includegraphics[scale=\figscale]{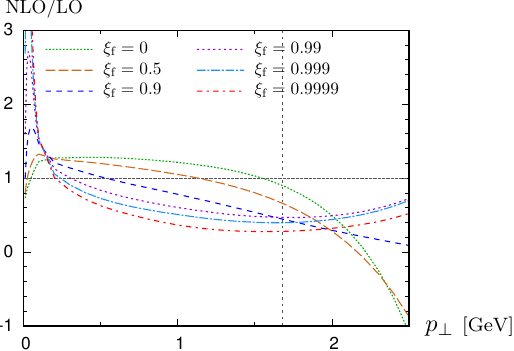}
	\caption{Left: Multiplicity as a function of $\ptt$ for different values of $\xif$. Right: Ratio of the multiplicity at NLO and LO for different values of $\xif$. In both cases the vertical dashed line corresponds to $\qs \approx \ptt$.}
	\label{fig:cutoff}
\end{figure}

We first consider a fixed value of the cutoff $\xif$. In Fig.~\ref{fig:cutoff}~(L) we show the multiplicity as a function of $\ptt$ for various values of $\xif$ between 0 and 1. When $\xif=0$ the multiplicity at NLO becomes negative for $\ptt$ values larger than about 2 GeV. This is similar to what was obtained in Ref.~\cite{Stasto:2013cha} in the same kinematics but considering all the channels. From the same figure we see that if we take $\xif$ close enough to 1 it is possible to make the multiplicity positive up to arbitrarily large values of $\ptt$. On the other hand, values of $\xif$ close to 1 lead to smaller multiplicities at small $\ptt$ as can be seen from Fig.~\ref{fig:cutoff}~(R).

\begin{figure}
	\centering
	\raisebox{1.5cm}{\includegraphics[scale=0.9]{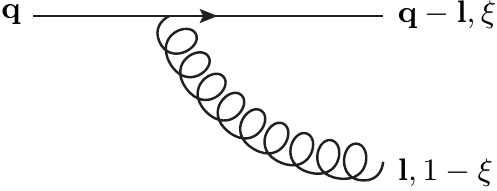}}
	\hspace{1.5cm}
	\includegraphics[scale=\figscale]{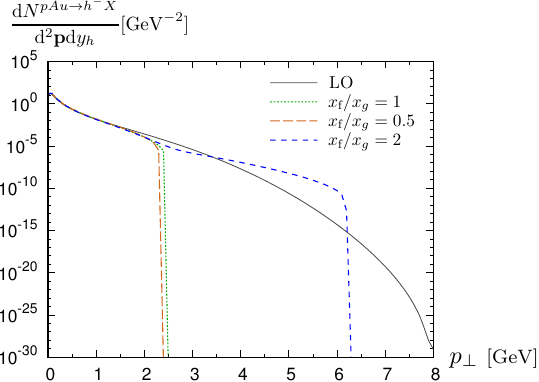}
	\caption{Left: Gluon emission. Right: Multiplicity obtained using different values of $\frac{\xf}{x_g}$.}
	\label{fig:xifcontinuous}
\end{figure}

Now we would like to fix $\xif$ to a reasonable value based on physical considerations. Let us consider the typical splitting diagram shown in Fig.~\ref{fig:xifcontinuous}~(L). The light cone energy $\Delta k^-$ needed from the target is
\begin{equation}\label{eq:kminusqg}
\Delta k^-_{qg} =\frac{x_g P^-}{\kt^2} \frac{(\lt-(1-\xi)\qt)^2}{\xi(1-\xi)}.
\end{equation}
We want to absorb fluctuations with a $\Delta k^-$ larger than a certain scale in the evolution of the target, i.e. contributions satisfying $\Delta k^-_{qg} \gtrsim \xf P^-$, where the natural value for the scale $\xf$ is of the order of $x_g$, the $k^-$ coming from the target at leading order. If all the transverse momenta involved are of similar magnitude, $\Delta k^-_{qg} \sim x_g/(1-\xi) \gtrsim \xf$ for all $\xi$ so one can take $\xif=0$ as in Refs.~\cite{Chirilli:2011km,Chirilli:2012jd}. On the other hand, if $\ktt$ is much larger than the saturation scale of the target $\qs$, the condition $\Delta k^-_{qg} \gtrsim \xf P^-$ is not always satisfied because of the integration over $\lt$ in a range involving values of $\ltt$ of the order of the saturation scale, $\ltt \sim \qs$. Therefore, in these kinematics, we should subtract contributions with $\xi$ close to 1 satisfying 
\begin{equation}
\Delta k^-_{qg} = 
\frac{x_g P^-}{\kt^2} \frac{\qs^2}{1-\xi} \geq \xf P^- \Leftrightarrow 1-\xi\leq \frac{\qs^2}{\kt^2} \frac{x_g}{\xf}. 
\end{equation}
To smoothly interpolate between the regions of small transverse momenta (where $\xif=0$) and large transverse momenta (where $\xif=1-\frac{x_g}{\xf} \frac{\qs^2}{\kt^2}$), we use $\xif(\ktt) = 1/(1+ \frac{x_g}{\xf}\frac{\qs^2}{\ktt^2})$.

In Fig.~\ref{fig:xifcontinuous}~(R) we show our results for the multiplicity with three different choices of $x_g/\xf$ between $\frac{1}{2}$ and 2. We observe that for these three values the multiplicity is negative above some $\ptt$. However the $\ptt$ value where this happens is very sensitive to the choice of $x_g/\xf$. In particular, the choice $x_g/\xf=0.5$, which is still in the ``natural'' range of this ratio, extends significantly the range of positivity of the multiplicity. We believe that this strong dependence of our results on the exact choice of this ratio comes from two aspects of our implementation that could be improved.
First, we tried to impose the condition $\Delta k^-_{qg} \gtrsim \xf P^-$ by using only external scales such as $\ktt$ and $\qs$. A more careful treatment would be required to impose this condition in an exact way in the transverse integrals of Eq.~(\ref{eq:J_Jv}). Second, we used the simple Golec-Biernat and Wüsthoff parametrization for the dipole cross section. In this model the LO term falls like a gaussian at large $\ptt$ while the NLO term has a power law behaviour. Consequently the sensitivity to the NLO corrections is very large. Using a dipole cross section obtained by solving the Balitsky-Kovchegov equation~\cite{Balitsky:1995ub,Kovchegov:1999yj} should reduce the importance of NLO corrections by making the LO term behaviour closer to a power law.

\section{Conclusions}

In this work we studied the effect of introducing an explicit rapidity factorization scale when subtracting the rapidity divergence in the NLO particle production cross section.
We have shown that it is possible to choose this scale so that the cross section is positive up to arbitrarily large transverse momenta. We have then suggested to use light cone energy ordering to fix this scale, but our final results are still very sensitive to variations of this scale in its ``natural'' range. Still, several improvements could be made to this work. First, one should impose light cone ordering in an exact way when performing the transverse momentum integrals. Second, one should apply this procedure with more physical dipole cross sections, such as one obtained by solving the Balitsky-Kovchegov equation.

\section*{Acknowledgements} 
We thank E. Iancu, Z. Kang, B.-W Xiao and D. Zaslavsky for discussions. This work has been supported by the Academy of Finland, projects 
267321 and 273464.

\providecommand{\href}[2]{#2}\begingroup\raggedright\endgroup

\end{document}